\documentclass[10pt, conference, doublecolumn]{IEEEtran}

\IEEEoverridecommandlockouts

\usepackage{config}

\begin{document}
\bstctlcite{IEEEexample:BSTcontrol}

\title{Reconfigurable Intelligent Surfaces: The New Frontier of Next G Security}

\author{
  \IEEEauthorblockN{
    Jacek Kibilda\IEEEauthorrefmark{1}, Nurul H. Mahmood\IEEEauthorrefmark{2}, Andr\'e Gomes\IEEEauthorrefmark{1}, Matti Latva-aho\IEEEauthorrefmark{2}, Luiz A. DaSilva\IEEEauthorrefmark{1} \\
  }
  \IEEEauthorblockA{
    \IEEEauthorrefmark{1}\textit{CCI, Virginia Tech}, USA,  e-mail: \{jkibilda,gomesa,ldasilva\}@vt.edu
  }
  \IEEEauthorblockA{
    \IEEEauthorrefmark{2}\textit{6G Flagship, CWC, University of Oulu}, Finland, e-mail: \{nurulhuda.mahmood,matti.latva-aho\}@oulu.fi
  }
}

\maketitle

\begin{abstract}
\Ac{RIS} is one of the significant technological advancements that will mark next-generation wireless. \ac{RIS} technology also opens up the possibility of new security threats, since the reflection of impinging signals can be used for malicious purposes. This article introduces the basic concept for a \ac{RIS}-assisted attack that re-uses the legitimate signal towards a malicious objective. Specific attacks are identified from this base scenario, and the \ac{RIS}-assisted signal cancellation attack is selected for evaluation as an attack that inherently exploits \ac{RIS} capabilities. The key takeaway from the evaluation is that an effective attack requires accurate channel information, a \ac{RIS} deployed in a favorable location (from the point of view of the attacker), and it disproportionately affects legitimate links that already suffer from reduced path loss. These observations motivate specific security solutions and recommendations for future work.
\end{abstract}


\IEEEpeerreviewmaketitle

\acresetall

\section{Introduction}


A \ac{RIS} is a planar surface comprising many reflective elements whose reflective coefficients are dynamically controlled and may be customized to the radio link \cite{yuan2021reconfigurable}. Passive \acp{RIS} do not amplify the reflected signals. Rather, by manipulating the amplitude and phase shifts of their reflective elements, they can: \begin{inparaenum}[(i)] 
\item arbitrarily scatter the impinged signal,
\item steer it in an arbitrary direction, or
\item re-shape it by concentrating or splitting its power in the desired direction(s). \end{inparaenum}
For instance, in the absence of an unobstructed direct path between the transmitter and the receiver, the scattered signals may be used to create an alternative path or simply produce controlled distortions to the channel in the form of multipath propagation.

This programmability of the wireless channel is a dual-use technology. On the one hand, it makes \ac{RIS} an attractive solution for improving the availability, resilience, and dependability of \ac{Next G}. On the other hand, it creates opportunities for launching new and potentially damaging over-the-air attacks, since the incident waves can also be scattered toward disrupting wireless links. The potentially low-cost design and low-power consumption of \acp{RIS} make it affordable to deploy a rogue device or compromise a legitimate device.

\begin{figure}[t]
    \centering
    \includegraphics[width=0.99\linewidth]{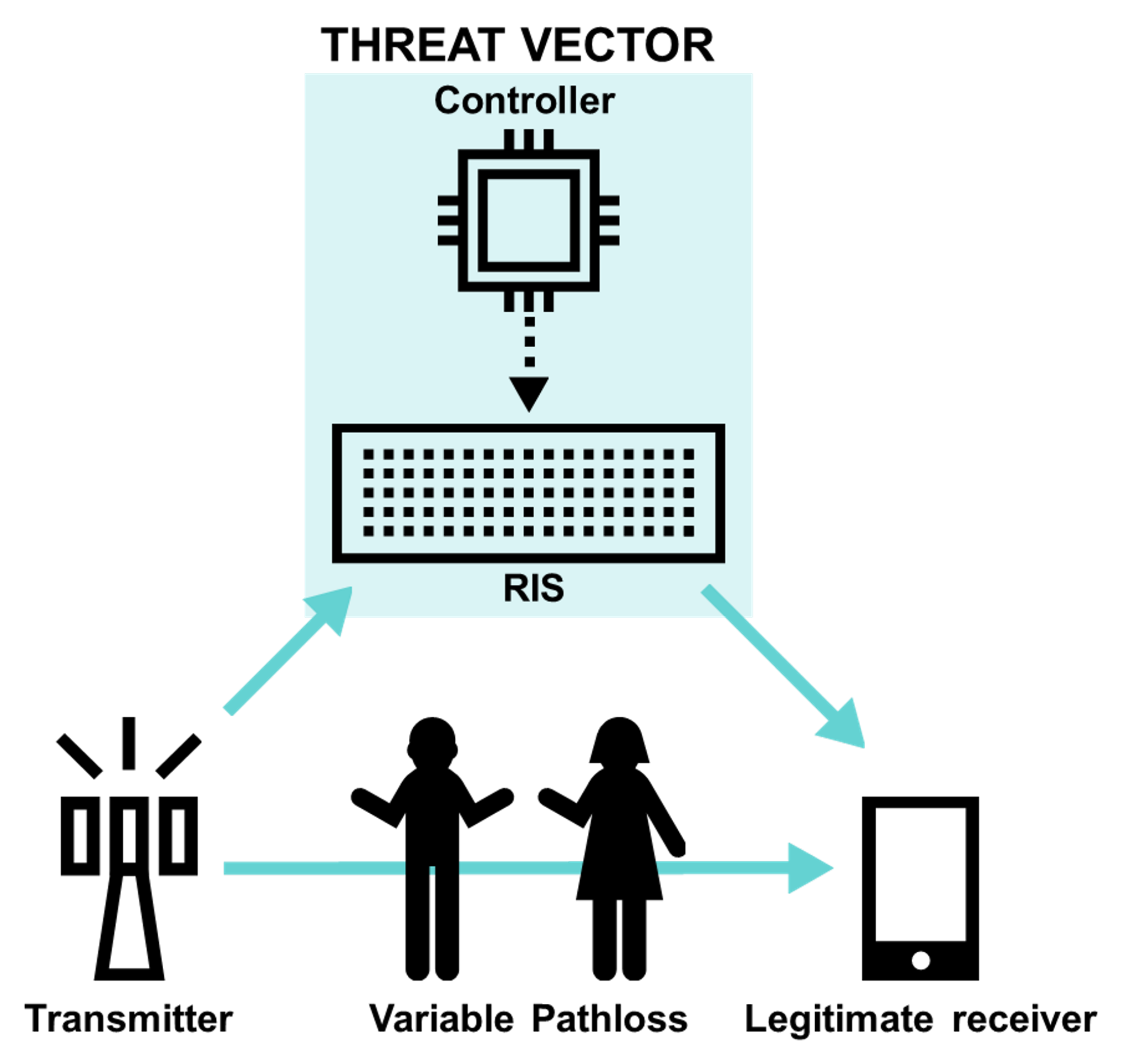}
    \caption{Illustrative scenario of a communication link under threat from an adversarial \ac{RIS}. 
    }
    \label{fig:model_attack}
\end{figure}

\Fig{model_attack} illustrates the basic concept of \iac{RIS}-assisted attack that re-uses the legitimate signal towards a malicious objective, exploiting wireless channel programmability and increasing the stealthiness of an adversarial act. Some already reported attacks create cancellation signals \cite{lyu2020irs}, disrupt the channel equalization \cite{staat2021mirror}, or poison the cell reference symbols \cite{yang2021novel}. 
The severity of these attacks ranges from reduced link quality to denial-of-service induced by the compromised wireless link initialization and control procedures. Despite this adversarial potential, there has been little work that reports on the fundamental challenges and effectiveness of triggering \ac{RIS}-enabled adversarial attacks.

This paper introduces a range of \ac{RIS}-assisted attacks, both previously reported in the literature and new, yet-to-be-analyzed attacks, and delves into the effectiveness of launching the most fundamental of these attacks the signal cancellation attack \cite{lyu2020irs}. This attack uses scattered propagation to create a cancellation signal that arrives at the receiver as destructive multipath, reducing the power of the legitimate link. Unlike \cite{wu2020towards,yuan2021reconfigurable,wang2022wireless}
, we take the perspective of an attacker to understand which characteristics make \acp{RIS} amenable for use in adversarial scenarios and how these characteristics may be used to counter such attacks. 

In what follows, we provide a brief background on \acp{RIS}. Then, following from the base scenario in \Fig{model_attack}, we identify and discuss different attacks and vulnerabilities that can be exploited using \iac{RIS}. From thereon, we focus on detailing the signal cancellation attack and quantifying its effectiveness. We consider how vulnerable a transmitter-receiver pair is to the attack given the influence of \ac{RIS} design, deployment location, and the accuracy of the channel estimates. Finally, we consider how the uncovered inefficiencies can be translated into secure network design and provide a set of prospective research areas that expand the scope of \ac{RIS}-assisted adversarial analysis.

\section{\ac{RIS} Characterization}
\label{sec:background}

\Iac{RIS} is a planar surface with programmable macroscopic physical characteristics whose electromagnetic wave response can be reconfigured by \iac{RIS} controller that either resides at the base station or is a standalone entity. \acp{RIS} can be of different types, the most common being the passive patch-array smart surfaces~\cite{LLM+21_RISprinciples}. This type of smart surface can be modeled as a collection of \ac{RIS} units. Each unit is equivalent to a lumped-element circuit with a tunable load impedance, and can be characterized by their reflection coefficients, modeled as $r_i = \beta_i \exp{(j \phi_i)},$ where $\beta_i$ and $\phi_i$ are the amplitude and phase responses, respectively. Passive \ac{RIS} elements only alter the direction of the reflected signal but do not amplify it. This is advantageous in that a passive \ac{RIS} consumes less energy compared to conventional relays.

However, the path loss of the wireless channel through the \ac{RIS} decays as the product of transmitter-\ac{RIS} and \ac{RIS}-receiver link path losses, making the equivalent channel rather weak~\cite{ZPR+22_activeRIS}. The double path loss attenuation also means that the performance of \iac{RIS}-assisted link depends on the \ac{RIS} position. It has been shown that this performance degrades as the \ac{RIS} moves further away from either the transmitter or the receiver, with the worst performance observed when it is equidistant from both~\cite{HAM+21_averageRate}. To address this limitation of passive \acs{RIS}, \textit{active \acp{RIS}} with an integrated active reflection-type amplifier supported by a power source have been introduced~\cite{ZPR+22_activeRIS}. This allows amplifying the amplitude response ($\beta_i$) along with adjusting the phase shifts ($\phi_i$) of each \ac{RIS} unit element, thereby enabling it to better focus the reflection towards specific directions.

Reaping the advantages of \ac{RIS} requires accurate \ac{CSI} on the channels to and from the \ac{RIS}. Since \iac{RIS} is typically composed of passive elements without any data processing capabilities, conventional \ac{CSI} estimation methods are not readily applicable. Instead, the channels must be estimated indirectly by the \ac{RIS} controller. The most common approach to do so involves using an unstructured model -- where the channels are simply described using complex gains -- and estimating the elements of the composite transmitter-\ac{RIS}-receiver channel. Such an approach is simple but incurs a considerable training overhead. Several methods for reducing the training overhead have been proposed, for example, based on grouping the \ac{RIS} elements, exploiting the common transmitter-\ac{RIS} channel, or parameterizing the channel by the angles of arrival, angles of departure, and the complex gains of each propagation path~\cite{SZL+22_channelEstimationRIS}. Any of the methods is likely to incur some error, which may potentially limit the \ac{RIS} performance.


Ideally, the phase shifts of the \ac{RIS} elements can take any value in the range $[0, 2\pi)$. In practice, the physical properties of the analog-to-digital converter limit the phase shifts to a discrete set of values whose cardinality scales with the number of quantization bits. The impact of the quantization error on the \ac{RIS} performance is not overwhelming when the number of quantization bits is two or more~\cite{HAM+21_averageRate}. This impact is further reduced when other non-idealities, such as \ac{CSI} estimation error~\cite{MBK+22_functionalArchitecture}, are considered.

\section{\ac{RIS}-assisted Attacks and Vulnerabilities}
\label{sec:summaryAttacks}

An adversary that compromises the \ac{RIS} controller or deploys its \ac{RIS} will be able to use the reflective properties of the \ac{RIS} to produce a targeted attack. In the following, we briefly introduce different \ac{RIS}-assisted attacks, some of which have already been reported in the literature while others are yet-to-be-analyzed as identifying \ac{RIS}-assisted attack vectors is an area of emerging research. \Tab{attacks} summarizes the identified attacks and provides a brief account of the exploitability and severity of each attack.

\begin{table*}[h]
 \centering
\scriptsize
\vspace{-0.1in}
\begin{tabular}{ l p{4cm} p{4cm} }
Attack & Exploitability & Severity \\
\hline
Signal Cancellation~\cite{lyu2020irs} & Requires high accuracy channel estimates \& solving complex optimization & It may dramatically reduce the amount of power at the receiver  \\
Pilot Sequence Poisoning~\cite{yang2021novel} &  Requires symbol-level synchronization and high accuracy channel estimates & Potentially may be used to re-direct the main link towards arbitrary location; in its less severe form, it may result in the selection of a sub-optimal beamforming vector\\
Beam Management Poisoning  & Requires timing synchronization with the network with beam sweeping and at least partial knowledge of channel estimates & Depending on the quality of synchronization and channel estimates, it may result in the selection of a sub-optimal beam and reduced link quality of even outage \\
Channel Equalization Poisoning~\cite{staat2021mirror} & Requires symbol-level synchronization and at least partial knowledge of channel estimates & May substantially degrade the quality of channel equalization resulting in low symbol error rate \\
PHY Key Generation~\cite{li2022reconfigurable} & Requires symbol-level synchronization with the main link and no or high accuracy channel estimates & Attack with random reflection coefficients reduces the secret key generation rate, but a more sophisticated attack may allow the attacker to introduce a backdoor by introducing arbitrary alteration in the secret key\\
PHY Authentication & Requires synchronization with the network and sub-symbol-level operation and no or high accuracy channel estimates & Simple attack may result in failed RF fingerprinting; advanced attack may allow the attacker to arbitrarily alter the identity of the node\\
\ac{RIS}-Aided Jamming~\cite{TSB22_EMI_RIS,wang2022wireless} & Requires active jammer \& partial knowledge of the jammer-\ac{RIS} channel & Reduces the \ac{SNR} at the receiver\\
\hline
\end{tabular}
  \caption{Exploitability and severity of \ac{RIS}-assisted attacks.}\label{tab:attacks}
 \vspace{-0.2cm}
 \end{table*} 

\subsubsection{Signal cancellation attack} Unlike a conventional jammer that aims to reduce the \ac{SINR} of the legitimate link by increasing the interference level at the receiver, a signal cancellation attack aims to create a signal that arrives at the receiver as a phase-reversed copy of the original signal that, at the receiver, will get destructively added to the original signal \cite{pan2020message}. Since passive \acp{RIS} do not modify the impinging signal they are ideally suited to launch a signal cancellation attack. An effective attack will require the reflection coefficients to be set such that the received power at the legitimate receiver is minimized. This requires channel estimates from all the links involved and solving a complex optimization problem akin to the one presented in \cite{lyu2020irs}.


\subsubsection{Attack against pilot sequences} Beamforming is critical to establishing coverage in ultra-high frequency systems that utilize large antenna arrays. In digital beamforming systems, appropriate beamforming vectors are estimated based on the \ac{CSI} estimated from transmitted pilot sequences. An adversarial \ac{RIS} may produce phase manipulations during the pilot sequence transmissions, resulting in a beam vector determined based on the manipulated \ac{CSI}. The attack is potentially trivial if the purpose is only to produce any sub-optimal beam vector, in which case applying random phase shifts during the pilot sequence transmission may be sufficient. On the other hand, forcing the transmitter to calculate an arbitrary beam vector will likely require high accuracy \ac{CSI} and solving a complex optimization problem akin to the system proposed in \cite{yang2021novel}, whereby the legitimate link is re-directed towards an eavesdropper.   

\subsubsection{Attack against beam management} In analog beamforming systems, the beam used for transmission comes from a set of pre-defined beam vectors that are periodically and iteratively used to sweep the cell area. The candidate beam used for transmission is the one for which the measured reference symbol received power is the strongest. In order to confuse the beam sweeping process, an adversarial \ac{RIS} may adjust its reflection coefficients to produce a scattered signal that will change depending on the beam being swept. The scattered signal, if directed at the receiver, may corrupt the received power measurements, resulting in the selection of a sub-optimal beam pair. This attack is easily triggered using an adversarial \ac{RIS}, as long as the \ac{RIS} knows the receiver's direction. The \ac{RIS} may re-direct the impinged signal towards the receiver, producing constructive or destructive multipath at the receiver. This multipath may reduce the received power on the best beam pair and increase it in other potential directions. The effectiveness of this attack will only increase if the \ac{RIS} can optimize its phase shifts based on full or partial \ac{CSI}.

\subsubsection{Attack against channel equalization} Accurate channel estimation is critical in equalizing the channel effects at the receiver. Suppose the channel varies significantly between the estimation and equalization phases. In that case, the result is poor quality decoding with a high symbol-error rate. \Iac{RIS} that can change reflection coefficients at the symbol rate may be used by an adversary to destroy channel coherence between channel estimation and equalization. The possibility of such an attack was described in \cite{staat2021mirror}. A practical attack requires an attacker to select sufficiently disruptive phase shift configurations. Moreover, faster than symbol-rate, phase shift alterations may produce power-leakage between subcarriers, further reducing the \ac{SNR}.

\subsubsection{Attack against \ac{PKG}} The reciprocity and uniqueness of a random wireless channel can be used to generate a shared secret key that two wireless nodes can use to encrypt their communication. A malicious \ac{RIS} may potentially trigger an attack that reduces the key generation rate by introducing random fluctuations to the channel at a rate higher than the channel sampling rate \cite{li2022reconfigurable}. This requires symbol-level synchronization with the main link. Suppose that, additionally, accurate \ac{CSI} is available to the adversarial \ac{RIS}. In that case, the attacker may try to optimize the reflection coefficients to introduce a backdoor into the shared key.

\subsubsection{Attack against \ac{PLA}} \Ac{PLA} is an emerging application where a wireless transmitter identity is tied to its location and channel-specific features that are discoverable by the receivers. An attacker \ac{RIS} may reduce the effectiveness of physical-layer-based authentication by producing fast enough random alterations to the phase shifts that will result in oscillations of the performance indicators used for authentication, such as the received signal strength or \ac{SNR}. This attack requires synchronization with the network and sub-symbol-level operation. If high accuracy channel estimates are available, it can potentially be used to arbitrarily disguise the identity of a node.

\subsubsection{RIS-aided jamming}
\Iac{RIS} may be used to aid an active jammer. The jammer transmits noise towards the \ac{RIS}, which reduces the effectiveness of the \ac{RIS} since the additional noise is \yale{relayed} together with the legitimate signal~\cite{TSB22_EMI_RIS,wang2022wireless}. The attack requires an active transmission towards the \ac{RIS} and can be targeted at any legitimate \ac{RIS}. In the presence of impinging noise, the \ac{SNR} of the legitimate link scales linearly with the size of the \ac{RIS}, as opposed to quadratic scaling when there is no attack.

\section{\ac{RIS}-Assisted Signal Cancellation Attack}



In this Section, we focus on quantifying the impact that \ac{RIS} design challenges, such as deployment location or accuracy of the channel estimates, have on the effectiveness and practicality of a signal cancellation attack. We selected this attack as our focus since it exploits capabilities inherent to \ac{RIS}, and launching it requires the ability to control these capabilities. In our evaluations, we consider the wireless system shown in \Fig{model_attack}. 


A legitimate transmitter communicates with a legitimate receiver in the presence of an adversarial \ac{RIS} deployed by a rogue actor or whose controller was compromised. The direct link between the transmitter and receiver is in either of two states, \ac{LoS} or \ac{NLoS}, depending on the absence or presence of an obstruction. The links to and from \ac{RIS} are assumed to be \ac{LoS}. The propagation is modeled using a bounded path loss model, with path loss exponents corresponding to the presence or absence of an obstruction. The \ac{RIS} is in the far-field of both transmitter and receiver, and the channel through the \ac{RIS} is cascaded. We assume that each individual channel is subject to independent Rayleigh fading. Rayleigh model represents here the worst-case fading. The \ac{LoS} and \ac{NLoS} path loss values, as well as fading models, follow the settings described in \cite{lyu2020irs}.

The transmitter is equipped with an 4x4 uniform planar array and steers its transmission using a beamforming vector corresponding to the receiver's direction. Meanwhile, the \ac{RIS} has $M$ reflecting elements whose adjustable reflection coefficients are optimized by the \ac{RIS} controller. Both the antenna design parameters and \ac{RIS} models correspond to the sub-\unit[6]{GHz} operation. The reception is assumed to be omnidirectional. The transmit power is \unit[20]{dBm} and noise floor is \unit[-90]{dBm}. The network topology is as follows: (a) transmitter at (0, 0); (b) receiver at (50, 0); and (c) \ac{RIS} at (45, 5).

We illustrate the effectiveness of the \ac{RIS}-assisted attack by observing the \ac{SNR} over one thousand realizations of the random channels. Solid lines trace the mean values observed, while shaded areas highlight the envelope corresponding to the 95\% confidence interval.

\subsection{\ac{RIS} Design}

In the first instance, we assume that the attacker can acquire exact and up-to-date information about all the channels and can use that information to optimize the phase shifts and magnitudes of reflection coefficients for each \ac{RIS} element (we will later relax this assumption). This joint optimization of \ac{RIS} phase shifts and magnitudes yields a mixed-integer non-linear optimization problem that is further complicated by the large number of reflective elements that need to be optimized. Generally, this class of problems is solved by a mixture of relaxations and iterative optimization, resulting in a heuristic solution. In this work, we apply a hierarchical approach with linear programming relaxations developed in \cite{lyu2020irs} that sequentially minimizes the received power in either the phase or magnitude domain, considering that phase and magnitude settings are conditionally independent.

\begin{figure}[t]
    \centering
    \includegraphics[width=0.99\linewidth]{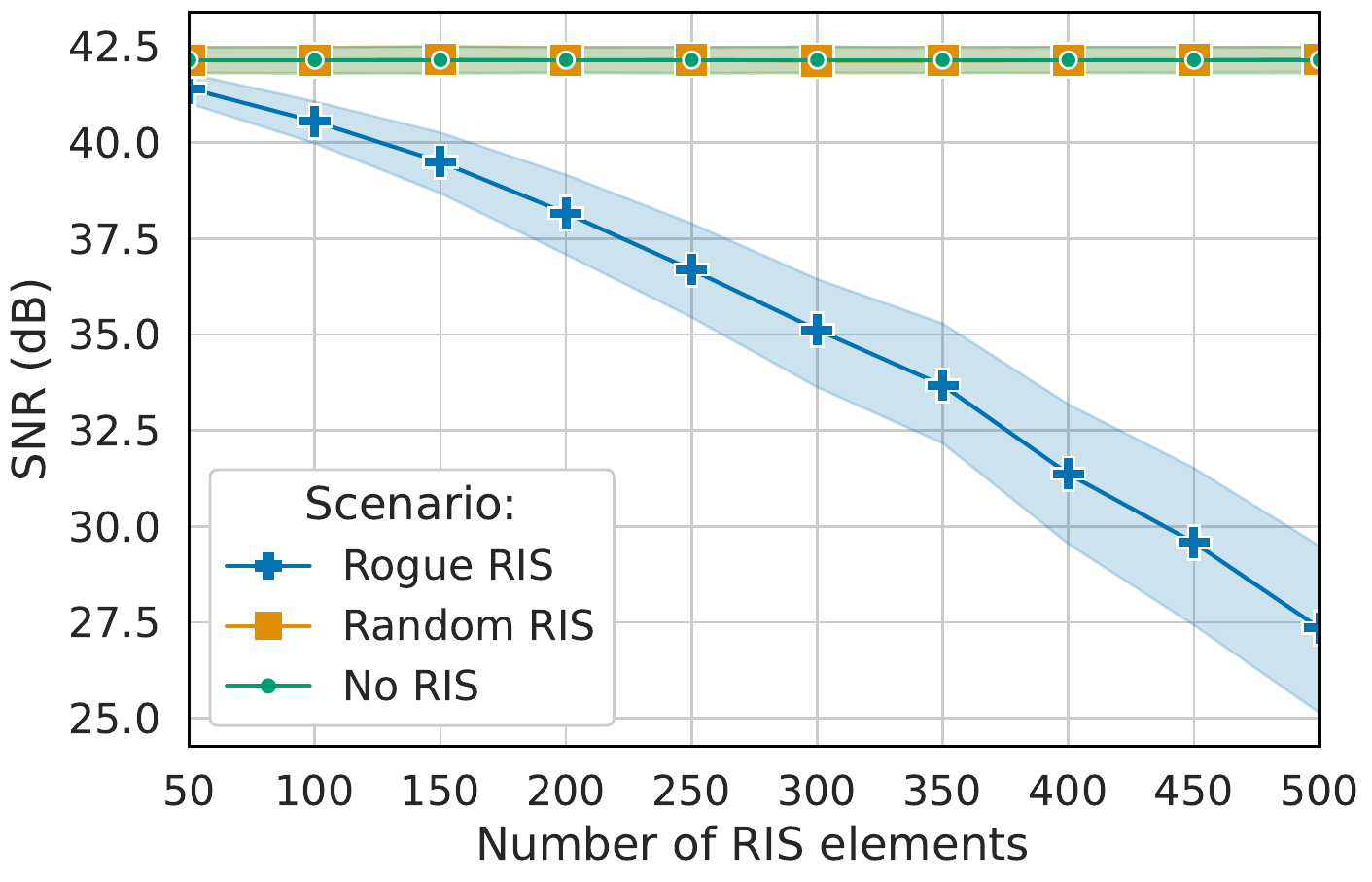}
    \caption{Effectiveness of the attack versus the number of \ac{RIS} elements in the \ac{LoS} case.}
    \label{fig:attack_ris_elems_los}
\end{figure}

\begin{figure}[t]
    \centering
    \includegraphics[width=0.99\linewidth]{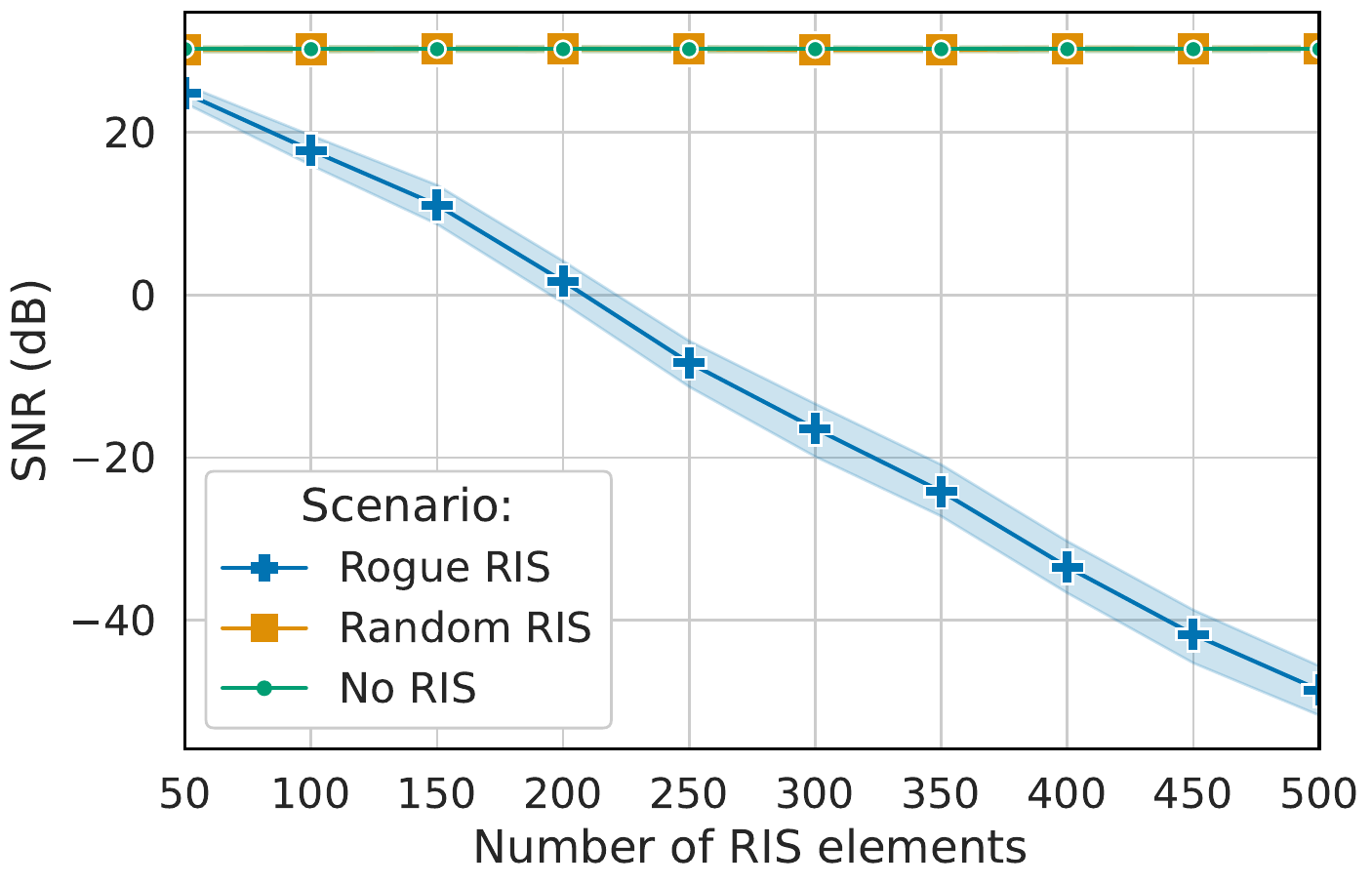}
    \caption{Effectiveness of the attack versus the number of RIS elements in the \ac{NLoS} case.}
    \label{fig:attack_ris_elems_nlos}
\end{figure}

This direct approach produces highly effective attacks, with the effectiveness of the attack scaling almost linearly with the size of the \ac{RIS}, as can be observed in \Fig{attack_ris_elems_los} and \Fig{attack_ris_elems_nlos}, corroborating the result presented in \cite{lyu2020irs}. Optimization of the reflection coefficients is an essential aspect of the attack since setting random phases on the \ac{RIS} produces negligible impact, equivalent to a scenario when the \ac{RIS} is inactive, denoted as \yale{no RIS}. Moreover, the attack requires a sizable \ac{RIS} to produce significant degradation of the \ac{SNR}. We see that the \ac{RIS} needs to have at least $50$ elements. Unsurprisingly, the attack is most effective when the direct path is obstructed. For \iac{RIS} of $450$ elements, we get, on average, almost \unit[12]{dB} reduction in link quality, for the \ac{LoS} case, but a likely outage in the \ac{NLoS} regime, courtesy to low \ac{SNR}. 



As discussed in \Sec{background}, practical \ac{RIS} designs offer only limited resolution phase shifts. From our observations, the resolution of the phase shifts has little impact on the effectiveness of the attack. We observed the reduction in the effectiveness of the attack only with the 1-bit resolution architecture (phase shifts of $0$ and $\pi$). This result corroborates observations made in \cite{HAM+21_averageRate} that transmission over \iac{RIS}, for large enough surface, is robust to even relatively low phase shift resolutions. Our model under observation is even more robust to quantized phase shifts since it involves a two-stage process in which the phase shifts are selected first. The reflection coefficient magnitudes are optimized after, potentially increasing the robustness of the attack.

\subsection{Relative Location}

\begin{figure}[t]
    \centering
    \includegraphics[width=0.99\linewidth]{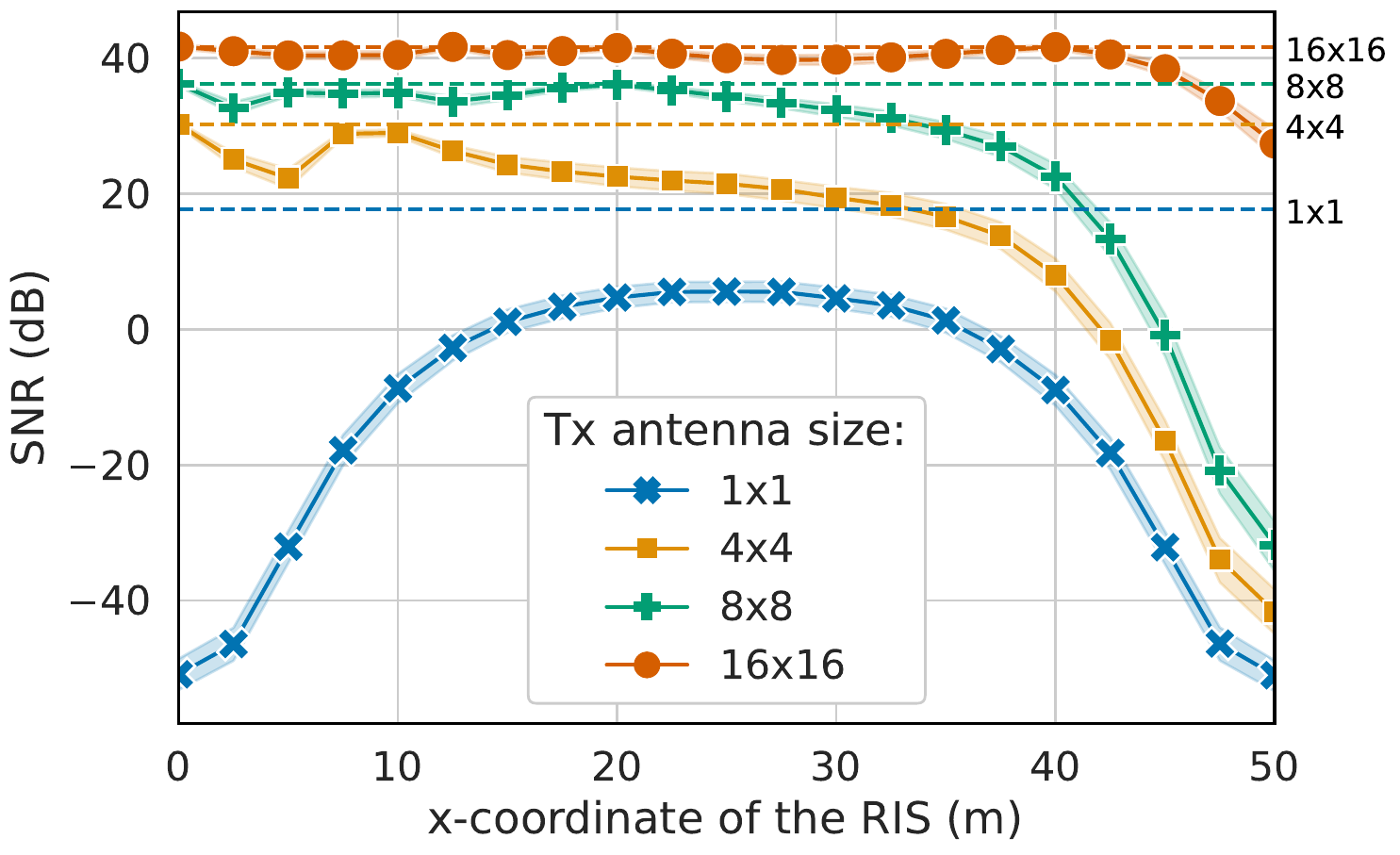}
    \caption{Effectiveness of the \ac{RIS}-assisted attack by a \ac{RIS} moving over a horizontal line connecting the points $(0,y)$ and $(50, y)$, with the transmitter and receiver located at $(0,0)$ and $(50, 0)$, respectively, and the direct link is \ac{NLoS}. Dashed lines act as reference for each transmitter array size setting, representing \ac{SNR} for a direct link with \yale{no \ac{RIS}}.}
    \label{fig:attack_dist_nlos}
\end{figure}

The performance of \iac{RIS}-assisted link depends on the relative position of the \ac{RIS}. Thus, in \Fig{attack_dist_nlos}, we observe how the link quality changes as we move the adversary \ac{RIS} between transmitter and receiver locations. Transmitter and receiver are located at coordinates $(0,0)$ and $(50, 0)$, respectively, whereas the \ac{RIS} is moved along the horizontal line connecting the points $(0,y)$ and $(50, y)$, whereby $y$ is set to \unit[5]{m}. The vertical distance is set arbitrarily to ensure far-field operation.

As we can observe, the attack is most effective (likely producing an outage) when the \ac{RIS} is close to the receiver. An attack triggered at a location far from the receiver produces a small effect (ca. \unit[5]{dB}) even in comparison with the direct link transmission with an inactive \ac{RIS}. One reason for this is that the signal scattered by the \ac{RIS} suffers significant path loss due to the cascaded nature of the channels, and this effect is strongest when the \ac{RIS} is at a point equidistant to the transmitter and receiver~\cite{HAM+21_averageRate}. We can observe this effect for a transmitter with omnidirectional radiation pattern.

However, in our scenario, the transmitter is equipped with an antenna array producing highly directional transmission. This means little power is emitted in the direction of an adversarial \ac{RIS}. Thus a \ac{RIS} located close to the transmitter cannot launch an effective cancellation attack. It is only when the \ac{RIS} is in the proximity of the receiver, capturing some of the energy radiated in the main-beam, when the attack becomes effective again. Generally, the more directional transmission, the less signal is irradiated towards the adversarial \ac{RIS}, weakening the potential attack. Increasing a vertical distance between the \ac{RIS} and receiver produces a similar effect. An attacker unable to position itself close enough to the receiver is required to have signal amplification capabilities \cite{ZPR+22_activeRIS}, making it more akin to an active jammer.


\subsection{Channel Estimate Error}

So far, we have assumed that an attacker has the accurate information required to conduct the attack. In particular, that means having exact knowledge of the three channels depicted in \Fig{model_attack} between the transmitter-receiver, transmitter-\ac{RIS} and \ac{RIS}-receiver. In practice, even a sophisticated adversarial \ac{RIS} that has either reception capabilities or access to multiple sensors will have only limited-accuracy channel estimates due to the added complexity of acquiring accurate \ac{CSI} for the channels associated with the \ac{RIS}, especially in the absence of active cooperation with either transmitter or receiver.

\begin{figure}[t]
    \centering
    \includegraphics[width=0.99\linewidth]{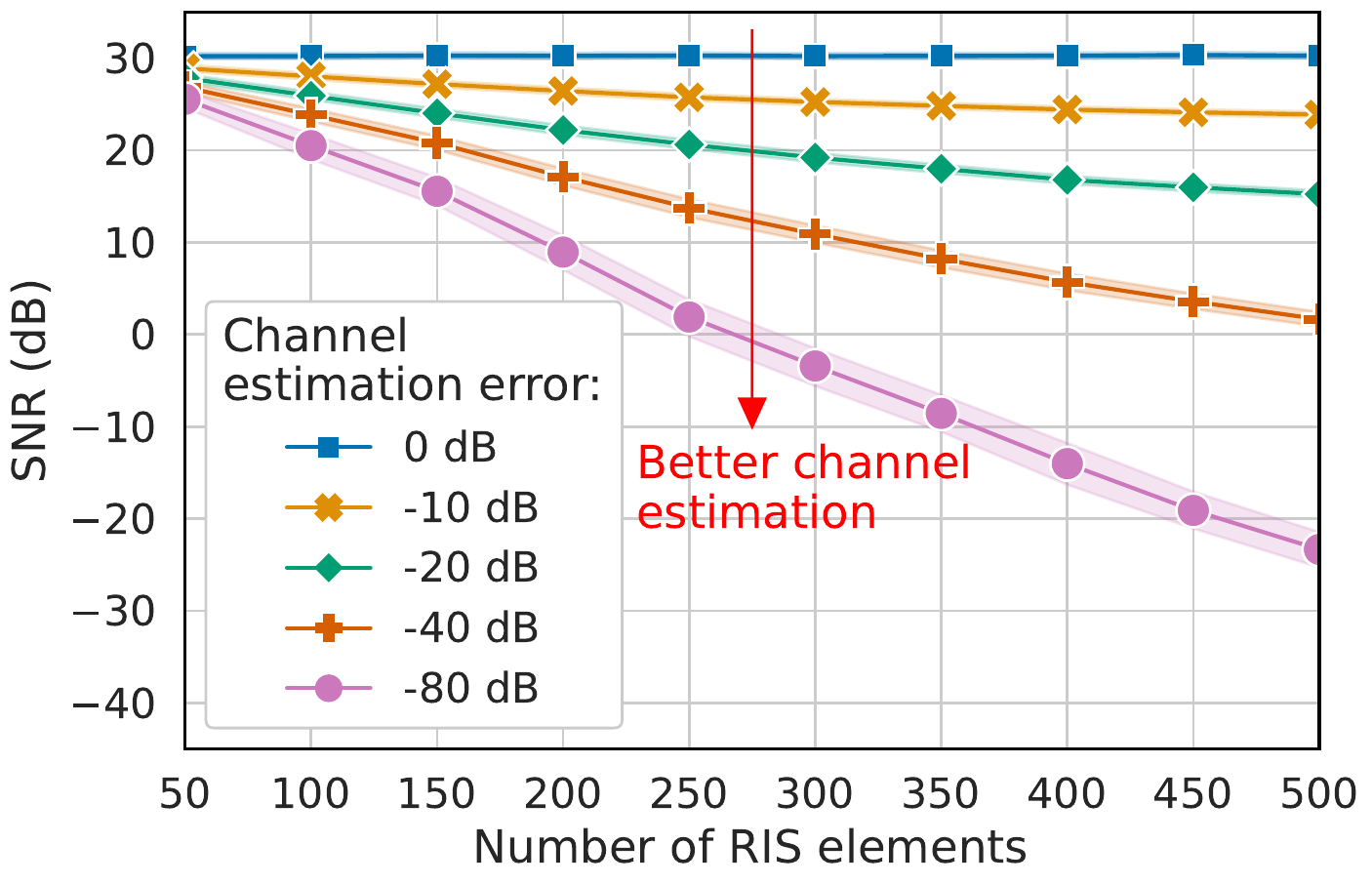}
    \caption{Effectiveness of the attack in the presence of channel estimate error. 
    }
    \label{fig:attack_mse_los}
\end{figure}

\Fig{attack_mse_los} illustrates the effectiveness of the attack when the channel estimates are subject to a given mean square error. The key observation here is that the attack loses much effectiveness in the presence of channel estimate errors. Noise in the channel estimates makes it difficult for the attacker to select the right combination of phases and magnitudes to ensure effective signal cancellation. Attacks with inaccurate channel state information yield a substantially weaker attack that may reduce the link quality, but is unlikely to cause an outage. It is only when the channel estimate precision is high enough (\unit[-80]{dB}) that for a reasonably sized \ac{RIS} we can observe an \ac{SNR} of \unit[0]{dB} or lower, potentially yielding an outage. An error of \unit[0]{dB} essentially corresponds to random coefficient selections. 

This result tells us that an effective attacker needs a method to optimize the \ac{RIS} reflection coefficients under \ac{CSI} uncertainty. Here, we conjecture that approaches that combine AI with sensing in the form of \acp{GAN} might be most effective from an attacker's point of view. We also observed that among the three channels involved in the scenario in \Fig{model_attack} the channel estimate on the direct link between the transmitter and receiver has the most impact on the effectiveness of the attack. Thus suggesting that the attacker needs to prioritize sample collection on this particular link.


\section{Implications to Next G Security and Future Directions} 

In this section, we extract key take-away messages and discuss their implications for the Next G network's security design. We also highlight future research directions for assessing the adversarial potential of \acp{RIS}.

\subsection{Key Take-away Lessons}

\subsubsection*{\ac{NLoS} users are most vulnerable} We have seen that the link quality degradation from \iac{RIS}-assisted attack is substantially higher for \ac{NLoS} receivers than the absolute difference between the path gains between the \ac{NLoS} and \ac{LoS} channels might indicate. This observation emphasizes the importance of \ac{LoS} transmission to resilient communication but also suggests that any security solutions, such as \ac{RIS} authentication, should primarily address vulnerable \ac{NLoS} links.
\subsubsection*{Attack effectiveness scales with the \ac{RIS} size} As is intuitively expected, the effectiveness of the attack increases with the size of the \ac{RIS}. This suggests that limiting the space available for deploying large-size \ac{RIS} is one strategy to reduce the effectiveness of any potential attack. Similarly, any legitimate \ac{RIS} comprised of panels should provide strong security for each individual panel, potentially limiting the effectiveness of hijacking attacks targeting a single panel. 
\subsubsection*{Attack is most effective when the \ac{RIS} is closer to the receiver} A \ac{RIS} placed far away from both transmitter and receiver produces negligible impact. But, equally, as the transmission becomes more directional the effectiveness of the attack is reduced. An effective safeguarding strategy would require ensuring that there is no space to deploy an illegitimate \ac{RIS} in the vicinity of receiver locations. Additionally, having in-built capabilities to localize and detect \acp{RIS} in the vicinity would help identify adversarial devices.
\subsubsection*{Adversarial \ac{RIS} requires high accuracy channel estimates} The effectiveness of the attack depends significantly on the accurate knowledge of the channels. Not all channels affect that effectiveness equally. In particular, the estimates of the direct link channel have a disproportionate impact on this effectiveness. This suggests that a particularly effective strategy to safeguard against the attack is to either ensure that the channel is sufficiently non-stationary or incoherent by including noise in the transmission. The alternative is to ensure appropriate capabilities to detect and eliminate malicious receivers and sensors that may help with accurate channel estimates.

\subsection{Outstanding Research Challenges in \ac{RIS} Security}

\subsubsection*{\ac{RIS}-powered
networks}

So far, we have considered a scenario involving a single uncoordinated \ac{RIS} allowing the re-configuration of a particular link. The concept can be extended toward an intelligent and reconfigurable wireless operating environment for a local area of the network~\cite{MBK+22_functionalArchitecture}. Such a paradigm integrates multiple spatially distributed \acp{RIS} within the local area to enable spatially focused service provisioning through coordinated optimization of the individual \ac{RIS} elements.

Despite its significant potential, a networked-\ac{RIS} enabled wireless environment also introduces novel multi-dimensional security risks. The number of vulnerable points increases in this case. An adversarial agent taking control of one or more \ac{RIS} within the network will have significant leverage over the overall network unless proper measures are taken. Launching jamming attacks on the different \acp{RIS} also becomes easier. Moreover, the multidimensionality of the network increases the risk of launching one or more of the possible attacks outlined in \Sec{summaryAttacks}.

\subsubsection*{\Ac{STAR}-\ac{RIS}}
Our discussions have been limited to a passive \ac{RIS}. Recent advancements in \ac{RIS} include the \ac{STAR}-\ac{RIS}~\cite{LMX+21_starRis}, also known as an intelligent omni-surface. 
By manipulating both the electric and magnetic currents of the \ac{RIS} elements, these \ac{RIS} variants allow a portion of the incident signal to be reflected (as in conventional passive \ac{RIS}) while the remainder is refracted (or otherwise transmitted). To the best of our knowledge, its physical layer security aspects are yet to be studied. Intuitively, the new architecture has the potential to significantly enhance security by refracting the desired signal while blocking the eavesdropped signal. 

\section*{Acknowledgements}
The research leading to this paper received support from the Commonwealth CyberInitiative (CCI), an investment in the advancement of cyber R\&D, innovation, and workforce development; and the Academy of Finland, 6G Flagship program under Grant $346208$. For more information about CCI, visit: \url{www.cyberinitiative.org.}. For more information about 6G Flagship, visit: \url{www.6gflagship.com}. We would also like to thank our colleagues Dr. Parth Pathak, Dr. Vijay Shah, and Dr. Kai Zeng for inspiring conversations that have led to formulating the list of potential adversarial attacks.

\bibliographystyle{IEEEtran}
\bibliography{references}

\begin{thebibliography}{10}
\providecommand{\url}[1]{#1}
\csname url@samestyle\endcsname
\providecommand{\newblock}{\relax}
\providecommand{\bibinfo}[2]{#2}
\providecommand{\BIBentrySTDinterwordspacing}{\spaceskip=0pt\relax}
\providecommand{\BIBentryALTinterwordstretchfactor}{4}
\providecommand{\BIBentryALTinterwordspacing}{\spaceskip=\fontdimen2\font plus
\BIBentryALTinterwordstretchfactor\fontdimen3\font minus
  \fontdimen4\font\relax}
\providecommand{\BIBforeignlanguage}[2]{{%
\expandafter\ifx\csname l@#1\endcsname\relax
\typeout{** WARNING: IEEEtran.bst: No hyphenation pattern has been}%
\typeout{** loaded for the language `#1'. Using the pattern for}%
\typeout{** the default language instead.}%
\else
\language=\csname l@#1\endcsname
\fi
#2}}
\providecommand{\BIBdecl}{\relax}
\BIBdecl

\bibitem{yuan2021reconfigurable}
X.~Yuan, Y.-J.~A. Zhang, Y.~Shi, W.~Yan, and H.~Liu,
  ``Reconfigurable-intelligent-surface empowered wireless communications:
  Challenges and opportunities,'' \emph{IEEE Wireless Communications}, vol.~28,
  no.~2, pp. 136--143, 2021.

\bibitem{lyu2020irs}
B.~Lyu, D.~T. Hoang, S.~Gong, D.~Niyato, and D.~I. Kim, ``{IRS}-based wireless
  jamming attacks: When jammers can attack without power,'' \emph{IEEE Wireless
  Communications Letters}, vol.~9, no.~10, pp. 1663--1667, 2020.

\bibitem{staat2021mirror}
P.~Staat, H.~Elders-Boll, C.~Zenger, and C.~Paar, ``Mirror mirror on the wall:
  Next-generation wireless jamming attacks based on software-controlled
  surfaces,'' \emph{arXiv e-prints}, pp. arXiv--2107, 2021.

\bibitem{yang2021novel}
J.~Yang, X.~Ji, F.~Wang, K.~Huang, and L.~Guo, ``A novel pilot spoofing scheme
  via intelligent reflecting surface based on statistical csi,'' \emph{IEEE
  Transactions on Vehicular Technology}, vol.~70, no.~12, pp. 12\,847--12\,857,
  2021.

\bibitem{wu2020towards}
Q.~Wu and R.~Zhang, ``Towards smart and reconfigurable environment: Intelligent
  reflecting surface aided wireless network,'' \emph{IEEE Communications
  Magazine}, vol.~58, no.~1, pp. 106--112, 2020.

\bibitem{wang2022wireless}
Y.~Wang, H.~Lu, D.~Zhao, Y.~Deng, and A.~Nallanathan, ``Wireless communication
  in the presence of illegal reconfigurable intelligent surface: Signal leakage
  and interference attack,'' \emph{IEEE Wireless Communications}, 2022.

\bibitem{LLM+21_RISprinciples}
Y.~Liu, X.~Liu, X.~Mu, T.~Hou, J.~Xu, M.~Di~Renzo, and N.~Al-Dhahir,
  ``Reconfigurable intelligent surfaces: Principles and opportunities,''
  \emph{IEEE Communications Surveys \& Tutorials}, vol.~23, no.~3, pp.
  1546--1577, 2021.

\bibitem{ZPR+22_activeRIS}
K.~Zhi, C.~Pan, H.~Ren, K.~K. Chai, and M.~Elkashlan, ``Active {RIS} versus
  passive {RIS}: Which is superior with the same power budget?'' \emph{IEEE
  Communications Letters}, vol.~26, no.~5, pp. 1150--1154, May 2022.

\bibitem{HAM+21_averageRate}
R.~Hashemi, S.~Ali, N.~H. Mahmood, and M.~Latva-aho, ``Average rate and error
  probability analysis in short packet communications over {RIS-Aided URLLC}
  systems,'' \emph{IEEE Transactions on Vehicular Technology}, vol.~70, no.~10,
  pp. 10\,320--10\,334, Oct. 2021.

\bibitem{SZL+22_channelEstimationRIS}
A.~L. Swindlehurst, G.~Zhou, R.~Liu, C.~Pan, and M.~Li, ``Channel estimation
  with reconfigurable intelligent surfaces--a general framework,''
  \emph{Proceedings of the IEEE}, pp. 1--27, May 2022.

\bibitem{MBK+22_functionalArchitecture}
N.~H. Mahmood, G.~Berardinelli, E.~J.~Khatib, R.~Hashemi, C.~de~Lima, and
  M.~Latva-aho, ``A functional architecture for {6G} special purpose industrial
  {IoT} networks,'' \emph{IEEE Transactions on Industrial Informatics}, 2022,
  early access.

\bibitem{li2022reconfigurable}
G.~Li, L.~Hu, P.~Staat, H.~Elders-Boll, C.~Zenger, C.~Paar, and A.~Hu,
  ``Reconfigurable intelligent surface for physical layer key generation:
  Constructive or destructive?'' \emph{IEEE Wireless Communications}, 2022.

\bibitem{TSB22_EMI_RIS}
A.~de~Jesus~Torres, L.~Sanguinetti, and E.~Bj\"ornson, ``Electromagnetic
  interference in {RIS}-aided communications,'' \emph{IEEE Wireless
  Communications Letters}, vol.~11, no.~4, pp. 668--672, Apr. 2022.

\bibitem{pan2020message}
Y.~Pan, Y.~Hou, M.~Li, R.~M. Gerdes, K.~Zeng, M.~A. Towfiq, and B.~A. Cetiner,
  ``Message integrity protection over wireless channel: Countering signal
  cancellation via channel randomization,'' \emph{IEEE Transactions on
  Dependable and Secure Computing}, vol.~17, no.~1, pp. 106--120, 2020.

\bibitem{LMX+21_starRis}
Y.~Liu, X.~Mu, J.~Xu, R.~Schober, Y.~Hao, H.~V. Poor, and L.~Hanzo, ``{STAR}:
  Simultaneous transmission and reflection for $360^\circ$ coverage by
  intelligent surfaces,'' \emph{IEEE Wireless Communications}, vol.~28, no.~6,
  pp. 102--109, Jun. 2021.

\end{thebibliography}

\end{document}